\documentclass[12pt,manuscript]{emulateapj}
\usepackage{graphicx}
\usepackage{amssymb}
\usepackage{amsmath}
\usepackage{natbib}

\newcommand{\lnm}{\bf LN}

\shorttitle{Statistical Cosmic Void Distribution}
\shortauthors{Pycke and Russell}

\begin{document}

\title{New Statistical Perspective to The Cosmic Void Distribution}


\author{J-R Pycke}
\affil{Division of Science and Mathematics, New York University Abu Dhabi, PO Box 129188, Abu Dhabi, UAE}
\email{jrp15@nyu.edu}

\and

\author{E. Russell}
\affil{Division of Science and Mathematics, New York University Abu Dhabi, PO Box 129188, Abu Dhabi, UAE}
\email{er111@nyu.edu}

\begin{abstract}
In this study, we obtain the size distribution of voids as a $3$-parameter redshift independent log-normal void probability function (VPF) directly from the Cosmic Void Catalog (CVC). Although many statistical models of void distributions are based on the counts in randomly placed cells, the log-normal VPF that we here obtain is independent of the shape of the voids due to the parameter-free void finder of the CVC. We use three void populations drawn from the CVC generated by the Halo Occupation Distribution (HOD) Mocks which are tuned to three mock SDSS samples to investigate the void distribution statistically and the effects of the environments on the size distribution. As a result, it is shown that void size distributions obtained from the HOD Mock samples are satisfied by the $3$-parameter log-normal distribution. In addition, we find that there may be a relation between hierarchical formation, skewness and kurtosis of the log-normal distribution for each catalog. We also show that the shape of the $3$- parameter distribution from the samples is strikingly similar to the galaxy log-normal mass distribution obtained from numerical studies. This similarity of void size and galaxy mass distributions may possibly indicate evidence of nonlinear mechanisms affecting both voids and galaxies, such as large scale accretion and tidal effects. Considering in this study all voids are generated by galaxy mocks and show hierarchical structures in different levels, it may be possible that the same nonlinear mechanisms of mass distribution affect the void size distribution.
\end{abstract}


\section{Introduction}

The seeds of the present-day large scale structure of the Universe are formed from random Gaussian density fluctuations in the early stages of its evolution. In these random density fluctuations, matter evolves from a linear to a highly nonlinear regime driven by gravitational instabilities. These instabilities in the primordial Gaussian density field form the complex structure of the universe that we observe today. As a result, it is natural to describe the probability distribution function of the density fluctuations as a crucial statistical tool to identify different types of environments of the Universe. For example, a Gaussian matter distribution function represents the linear regime, while the distribution function that deviates from Gaussian shows a highly nonlinear regime. In this framework, there are two important building blocks of the large scale structure; galaxies/overdense regions and voids/underdense regions. Initially these two features are formed from the same primordial Gaussian density field. While voids are formed from the minima of the density field, galaxies are formed in the density maxima. In time when galaxies grow in mass, voids tend to empty in their mass due to their peculiar gravitational force through the nonlinear regime. Although there are different phenomenological models of the probability distribution function (PDF) of overdense regions (such as galaxies, galaxy clusters,..etc.) in nonlinear regimes \citep{saslaw,Lahav93,Gaztanaga93,Ueda}, the statistical models of void probability functions (VPFs) \citep{fry86Voids,elizaldegaztanaga92,colesjones91} are based on the counts in randomly placed cells following the prescription of \cite{white79}. In addition to this VPF, the number density of voids is another key statistic to obtain the void distribution. \cite{patiri2006a} show that this number density can be estimated analytically by using numerical simulations or mock catalogues.

The phenomenological models of PDFs of overdense regions, especially the one-point and the two-point PDFs, are well studied. The two-point PDF is a tool to model the dark halo biasing as well as obtaining the errors in the one-point statistics \citep{ColombiBS1995,SzapudiC1996} while the one-point PDF is a useful tool to show the clustering of the Universe with higher order moments-statistics such as skewness and kurtosis \citep{Kayo}. Observations as well as models based on cold dark matter numerical simulations of galaxy distributions indicate that the density distribution is well approximated by a log-normal rather than a Gaussian one \citep{Hamilton1985,Bouchet1993,colesjones91,Kofman,TaylorWatts2000}. Also, \cite{Bernardeau92,Bernardeau94} show that the PDF computed from perturbation theory in a weakly nonlinear regime approaches log-normal when the primordial power spectrum is proportional to index $n=-1$. Later on, \cite{Kayo} find the one-point log-normal PDF can describe the density distributions accurately, not only in weakly nonlinear, but also in the highly nonlinear regime by using the nonlinear density fluctuations from $N$-body simulations with Gaussian initial conditions. \cite{Kayo} also indicate that this PDF is fairly independent of the shape and the power spectrum of density fluctuations. However the underlying mechanism of the cosmological origin of the log-normal distribution remains unknown.

In this study we determine that a specific standard parametric model can fit samples of void radii from the Cosmic Void Catalog (CVC) of \cite{sutter} after suitable estimation of the parameters in the model. The three mock catalogs under study, referred to by the nicknames given in the CVC, are N-body Mock, HOD Sparse and HOD Dense. Using these samples we show that the system of $3$-parameter log-normal distribution provides a fairly satisfactory model of the size distribution of voids which is strikingly similar to the galaxy mass distribution of \cite{Kayo}, assuming the void size is proportional to its mass content ($R^3\approx M$). To do this, we follow three standard steps in the statistical analysis. These are the exploratory phase, the choice of a probability model, estimations of the parameters of the model and finally the graphical and numerical assessment (see Chapter 4 \S Section 4.1 and 4.2 in \cite{fisher} and the Introduction in \cite{sheskin} for an overview of basic principles and some terminology employed in the fields of descriptive statistics, estimation and goodness-of-fit procedures).

\section{Void Catalog and Numerical Data}

To investigate the void size distribution function statistically in simulations, here we use the Public CVC of \cite{sutter}. CVC has two main catalogs: a complete catalog that is suitable for void galaxy surveys, and a bias-free catalog of voids that provides a fair sampling of void shapes and alignments in which the void effective radii vary between $5$ and $135$ $Mpc/h$ \citep{sutter}. All the data samples we use here from CVC are generated from a $\Lambda$ cold dark matter ($\Lambda$CDM) N-body simulation by using an adaptive treecode N-body method called the 2HOT code \citep{sutter2014a,sutter2014b}. \cite{sutter2014a,sutter2014b} extract halos from the N-body simulation, and use the position and masses of halos to produce a HOD model. Therefore, galaxy catalogues are produced from a halo population by using the Halo Occupation Distribution (HOD) code of \cite{Tinker2006} and the HOD model by \cite{Zheng07}. \cite{sutter2014a} and \cite{sutter2014b} generate three mock catalogs; HOD Dense, HOD Sparse and N-body Mock. In these catalogs, voids are identified with by the modified version of the parameter-free void finder ZOBOV \citep{Neyrinck2008,Lavaux12,sutter}.

HOD Dense and HOD Sparse are produced by dark matter simulations of $1024^3$ particles in a $1$ Gpc/h box and all particles are kept in real space at $z=0$ and are tuned to the observational HOD fits \citep{sutter2014a}. As a result, the difference between these two catalogs comes from their resolution as well as their tuned observational data sets. HOD Sparse mock catalog consists of $1422$ voids with $14$ $Mpc/h$ effective minimum radii ($R_{eff,min}= 14$ $Mpc/h$) and this void catalog represents a relatively low resolution galaxy sample with density $3\times 10^{-4}$ particles per cubic $Mpc/h$ matching the number density and clustering of the SDSS DR$9$ galaxy sample \citep{Dawson2013} using the parameters found by \cite{Manera2013} ($\sigma_{\log M}=0.596$, $M_{0}=1.2\times10^{13}$ $h^{-1}M_{\odot}$, $M^{\prime}_{1}=10^{14} h^{-1}M_{\odot}$, $\alpha=1.0127$, and $M_{min}$ chosen to fit the mean number density). HOD Dense catalog has $9503$ voids with effective minimum radii $R_{eff,min}= 7$ Mpc/h and includes relatively high-resolution galaxy samples with density $4\times10^{-3}$ dark matter particles per cubic Mpc/h matching the SDSS DR$7$ main sample \citep{Strauss2002} using one set of parameters found by \cite{Zehavi2011} ($\sigma_{log M}=0.21$, $M_{0}=6.7\times10^{11} h^{-1}M_{\odot}$, $M^{\prime}_{1}=2.8\times10^{13}h^{-1}M_{\odot}$, $\alpha=1.12$). N-body Mock catalog is a single HOD Mock in real space at $z=0.53$, generated by a dark matter simulation of $4096^3$ particles (with a particle mass resolution $7.36\times 10^{10}$ $h^{-1}M_{\odot})$ in a $4$ Gpc/h box and is tuned to SDSS DR$9$ in full cubic volume by using the HOD parameters found in \cite{Manera2013}. Although N-body Mock catalog is processed slightly differently than HOD Sparse and HOD Dense, it is a HOD mock catalog and it uses Planck first- year cosmological parameters \citep{PlanckCollaboration}. The N-body Mock consists of $155,196$ voids \citep{sutter2014b}.

It is particularly important to mention that both the N-body Mock and HOD Sparse, represent voids generated from the relative low density galaxy mocks, while HOD Dense consists of void populations generated from a relatively high density galaxy mock catalog. As a result, these samples may help us to identify some possible relations between the void size distribution and the environment. Here we take into account maximum tree depth as an indicator of an environment. Note that the maximum tree depth is the length from root to tip of the tallest tree in the hierarchy as is defined by \cite{sutter2014a}, and it shows the amount of substructures in the most complex void \citep{sutter2014a}. According to this, the HOD Dense and Sparse have the maximum three depths $10$ and $4$, respectively \citep{sutter2014a}, while N-body Mock shows only the root voids, at the base of the tree hierarchy, and therefore they do not have parents which indicates that the maximum three depth of this sample $0$ \citep{sutter2014b}.

Here the samples HOD Sparse, HOD Dense and N-body Mock are denoted as $R^{(1)}=(r^{(1)}_i)_{1\leq i\leq 1,422}$, $R^{(2)}=(r^{(2)}_i)_{1\leq i\leq 9,503}$ and  $R^{(3)}=(r^{(3)}_i)_{1\leq i\leq 155,196}$ respectively. While $R^{(i)}$ ($i= 1, 2, 3$) represent the names of the samples, $r^{(i)}$ stand for data points.

\section{Exploratory phase}

The first step in the exploratory phase consists of examining the basic graphical and numerical indicators of the samples as histograms, parameters of location (range, mean, median and mode) or dispersion (standard deviation) and shape (skewness, kurtosis).

Let us first consider the raw data plots. Informative features appear from a cursory inspection of the histograms. Whereas those of $R^{(1)}$ and $R^{(2)}$ (see Figure 3) present the features of samples drawn from a single parent population, an unexpected local mode occurs in $R^{(3)}$ about the value $50$ Mpc/h, see Figure 1 Left Panel. This suggests that N-body Mock, $R^{(3)}$, should be usefully considered as a sample drawn not from a single population but from two or more void populations with different central density values. In other words it is reasonable to perform a task of classification. The latter is meant in the classical sense in mathematical statistics, as, e.g.,  in \S{44.1} of \cite{stuart3}); it consists of differentiating between two or more populations on the basis of multivariate measurements.

\begin{figure*}
\hspace{-10mm}
\begin{tabular}{ll}
\includegraphics[scale=0.55]{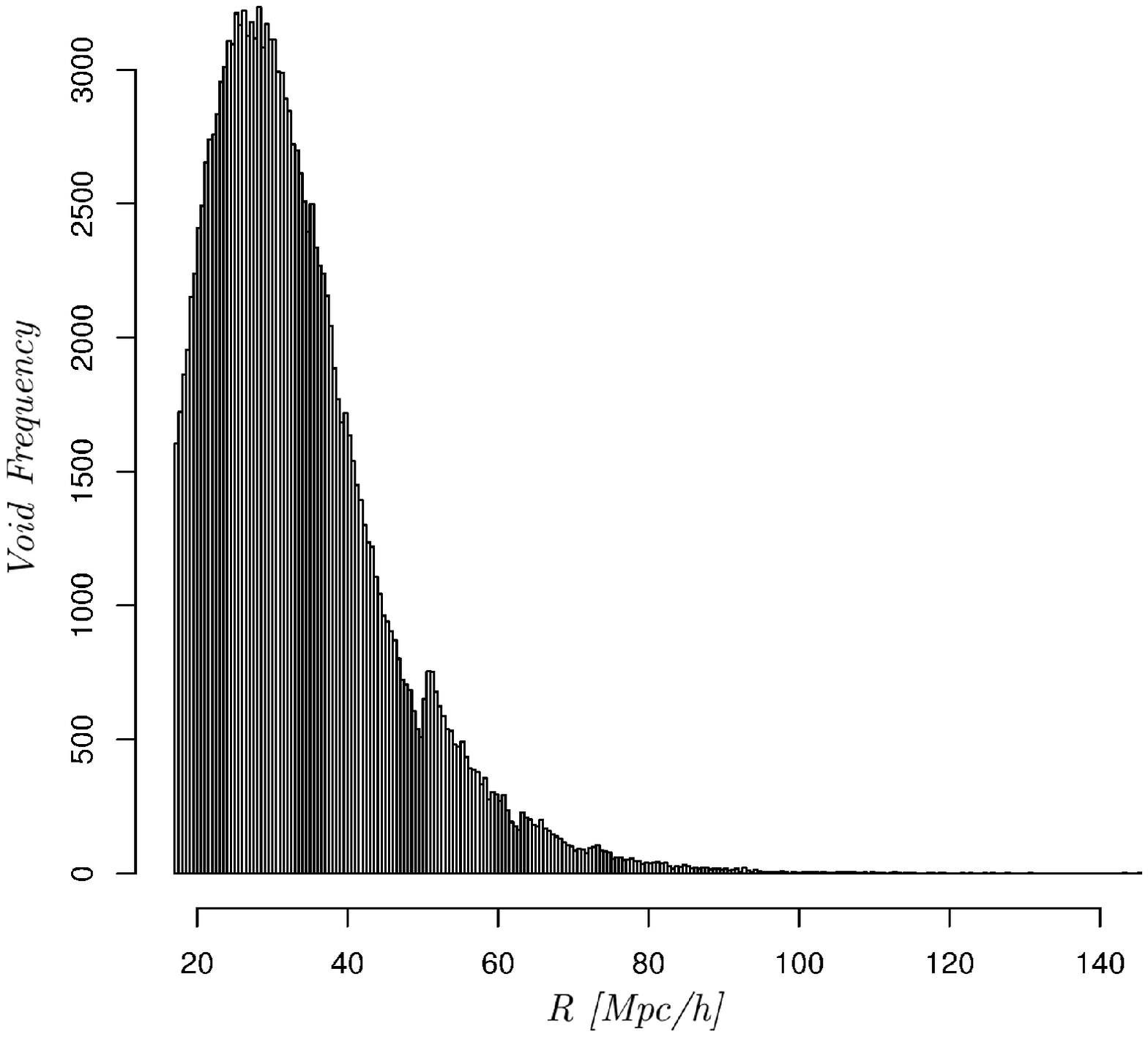}
\includegraphics[scale=0.55]{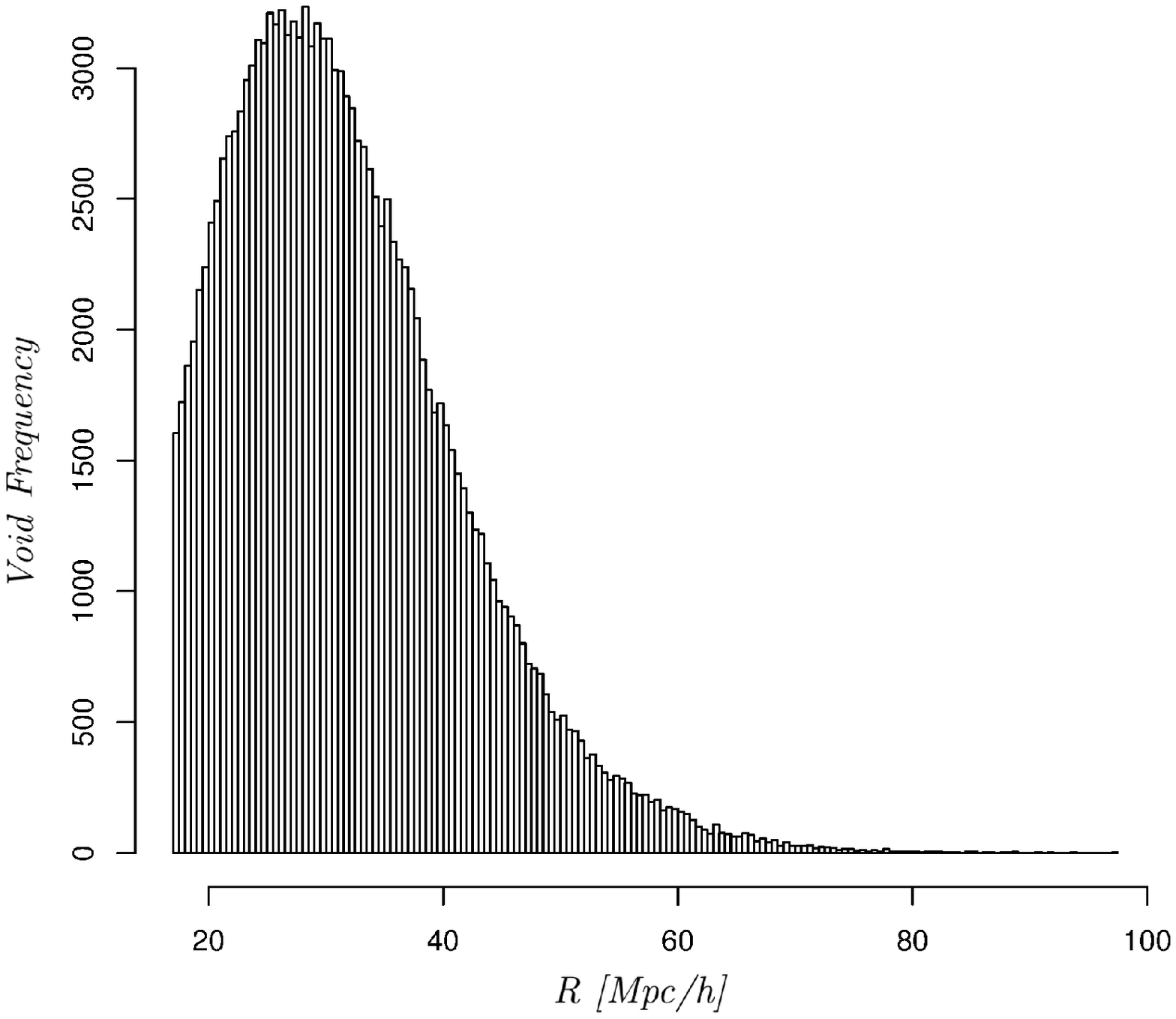}
\end{tabular}
\caption{Void size distributions of full N-body Mock data set, $R^{(3)}$ (Left Panel) and its sub-sample, $R^{(3)}_{0}$ consisting of the voids with only zero central density (Right Panel).}
\end{figure*}
\noindent
In this respect a closer inspection of the fourteen remaining numerical characteristics of the voids suggests that the central density will provide us with a criterion of classification. Out of the $155,196$ voids under study, $147,528$ voids have a  central density equal to zero. The sub-sample of $R^{(3)}$ consisting of the radii of these voids will be denoted $R^{(3)}_0$. Remarkably, the local mode observed in $R^{(3)}$ does not occur any more in $R^{(3)}_0$. The latter satisfies the basic graphical properties expected from a sample drawn from a single population, see Figure 1 (Right Panel).

Let us now turn to the numerical study of our samples $R^{(1)}$, $R^{(2)}$ and $R^{(3)}_0$.
The mean and the centered moments of  a sample $R=(r_i)_{1\leq i\leq N}$ are  $\overline{R}=N^{-1}\sum_{i=1}^{N} r_i$ and $m_k=N^{-1}\sum_{i=1}^N(r_i-\overline  R)^k, k=2,3,...$. The variance  $m_2$ is the most common index  of dispersion whereas widely used indices of the shape of a sample distribution are the  skewness  $b_1=m_3^2/m_2^3$ and  kurtosis $b_2=m_4/m_2^2$ (see formulas $(1.235)-(1.236)$ p. 51 in \cite{kotz}, $(3.85)-(3.86)$ p. 85 in \cite{stuart1}). The values of these moments computed from the samples are given in Table 1.

\begin{table}
\caption{First moments, skewness $b_1$ and kurtosis $b_2$ of  the sample distributions.}
$$\begin{array}{|c|c|c|c|}\text{Sample}&R^{(1)}&R^{(2)}&R^{(3)}_{0}\\\hline
\overline{r}&40.415\ &16.671\ &31.978\\
m_2&235.752\ &40.052\ &96.034\\
m_3&4,416\ &386.278\ &898.965\\
m_4&306,994&11,109\  &38,131 \\
b_1&1.488 &2.322\ &0.912\\
b_2&5.523\ &6.925 &4.134\\
\end{array}$$
\end{table}
\noindent
In Table 1, we see that the three empirical distributions share the property of being significantly positively skewed ($b_1>0$) and leptokurtic ($b_2>3$). Recall that the skewness $b_1$ measures the degree to which a distribution is asymmetrical. The kurtosis $b_2$ measures the curvature of a distribution. While a leptokurtic (resp. platykurtic) distribution is characterized by a high (resp. low) degree of peakedness, a mesokurtic distribution presents a peakedness considered medium whenever $b_2=3$, which is the kurtosis of a normal distribution with arbitrary average and variance (see \cite{sheskin} p. 16-30 for more details). Among the classical parametric families of distributions, the density of the three-parameter log-normal random variable $\mathbf R=\lnm(\theta,\zeta,\sigma)$ given by,

\begin{equation}
p_{\theta,\zeta,\sigma}(r)=\frac{e^{-\frac{\left\{\log(r-\theta)-\zeta\right\}^2}{2\sigma^2}}}{(r-\theta)\sigma\sqrt{2\pi}},\quad r>\theta,
\label{eqn:void-distribution}
\end{equation}
(see $(14.2)'$ p.~208 in \cite{kotz}) appears as a natural candidate to fit the size distributions of the samples HOD Dense, HOD Sparse and N-body Mock. A random variable ${\mathbf R}$ follows the distribution of $\lnm(\theta,\zeta,\sigma)$ if $\log(\mathbf R-\theta)$ follows a Gaussian distribution with mean $\zeta$ and variance $\sigma^2$. This three-parameter log-normal distribution is strikingly similar to the galaxy distributions that \cite{Kayo} obtain from the $N$-body simulations. Our motive is confirmed by the close proximity of the pairs $(b_1, b_2)$, computed from our sample, to the log-normal line observed in the skewness-kurtosis plane represented in Figure 2 (see Figure~12.3 in \cite{kotz} for a similar pattern involving several parametric families of distributions). As is seen in Figure 2, the void distributions from the mock samples of CVC can be considered to behave as log-normal distributions with respect to their skewness and kurtosis.

\begin{figure}
\centering
\includegraphics[scale=1.2]{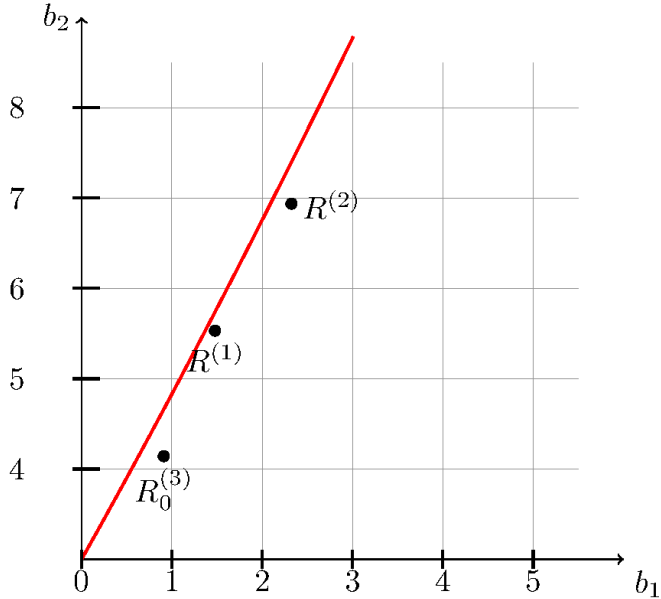}
\caption{Kurtosis $b_2$ as ordinate against the skewness $b_1$ as abscissa for the log-normal distribution (red solid line). The dots represent the kurtosis and the skewness of the size distributions of voids for HOD Sparse $R^{(1)}$, HOD Dense $R^{(2)}$, and the sub-sample of N-body Mock $R^{(3)}_{0}$.}
\end{figure}
\noindent
Some characteristics of a random variable $\lnm(\theta,\zeta,\sigma)$ are,

\begin{enumerate}
  \item range: $(\theta,\infty)$,
  \item mode: $\theta+e^{\zeta-\sigma^2}$,
  \item median: $\theta+\exp\zeta$,
  \item mean: $\theta+e^{\zeta+\sigma^2/2}$.
\end{enumerate}
This concludes the exploratory phase of our study and we will now deal with  the task of estimating the three parameters of the log-normal model for $R^{(1)}$, $R^{(2)}$ and $R^{(3)}_{0}$.

\section{Estimation}
The task of  estimation for a $3$-parameter log-normal distribution involves well-known computational difficulties related to the non-convergence of the maximum-likelihood estimator for the threshold parameter $\theta$. This issue is discussed in \cite{kotz} \S{4.2}. Therefore the most convenient standard tool is the moment-method. This is consists of equating the first three sample moments $\overline{R}$, $m_2$, and $m_3$ to the corresponding population values. This estimation method is described in \cite{kotz} \S{4.2}, in particular formulas $(14.45)-(14-46)$ p. 228. It proceeds as follows. First, one has to compute the unique real-valued solution $\tilde \omega$ of the equation,

\begin{equation}
\left(\tilde \omega-1\right)\left(\tilde \omega+2\right)^2=b_1=m_3^2/m_2^3,
\end{equation}
which is easily seen to be  given by the explicit formula,

\begin{eqnarray}\tilde{\omega}&=&\left[1+b_{1}/2+\sqrt{(1+b_{1}/2)^2-1}\right]^{1/3}\nonumber\\
&+&\left[1+b_{1}/2-\sqrt{(1+b_{1}/2)^2-1}\right]^{1/3}-1,
\label{eqn:distribution-parameters}
\end{eqnarray}
(see \cite{kotz}, last equality p.~228). Once this auxiliary coefficient has been determined, the estimators $(\hat\theta, \hat\zeta, \hat\sigma)$ of $(\theta,\zeta,\sigma)$ are,

\begin{enumerate}
  \item $\hat{\sigma}=\log\left({\tilde{\omega}}\right)^{1/2}$,
  \item $\hat{\zeta}=\frac 12 \log [m_2\tilde \omega^{-1}(\tilde \omega-1)^{-1}]$,
  \item $\hat{\theta}=\overline{R}-\tilde \omega^{1/2}e^{\hat \zeta}$,
\end{enumerate}
(see formulas p. 227-228 of \S{4.2} in \cite{kotz}). The estimates computed from our samples are given in Table 2.

\begin{table}
\caption{Estimates computed from the sample distributions.}
$$\begin{array}{l|r|r|r}\text{sample}&R^{(1)}&R^{(2)}&R^{(3)}_{0}\\\hline
\hat \theta&0.770&3.284&0.223\\
\hat \zeta&3.610&2.493&3.412\\
\hat \sigma&0.373&0.449&0.301\\
\end{array}
$$
\end{table}
\noindent
The goodness-of-fit of our model is illustrated in two ways. First, numerically by Table 3, Table 4, and Table 5 where expected values and observed values are given for three void distributions.

\begin{table}
\caption{Interval, observed and expected void frequencies of the data set $R^{(1)}$ (HOD Sparse), $\lnm(0.770, 3.610, 0.373)$}
$$\begin{array}{l|r|r|r}\text{Interval}&\text{Observed\phantom{a}Frequencies} & \text{Expected\phantom{a}Frequencies}\\\hline
$[0,10)$&0&0\\
$[10,20)$&56&57\\
$[20,30)$&334&320\\
$[30,40)$&429&424\\
$[40,50)$&296&306\\
$[50,60)$&163&168\\
$[60,70)$&77&81\\
$[70,80)$&37&37\\
$[80,90)$&16&16\\
\geq 90&14&13
\end{array}
$$
\end{table}

\begin{table}
\caption{Interval, observed and expected void frequencies of the data set $R^{(2)}$ (HOD Dense), $\lnm(3.284, 2.493, 0.449)$}
$$\begin{array}{l|r|r|r}\text{Interval}&\text{Observed\phantom{a}Frequencies} & \text{Expected\phantom{a}Frequencies}\\\hline
$[0,5)$&0&0\\
$[5,10)$&953&901\\
$[10,15)$&3448&3576\\
$[15,20)$&2947&2782\\
$[20,25)$&1267&1326\\
$[25,30)$&500&547\\
$[30,35)$&215&218\\
$[35,40)$&99&88\\
$[40,45)$&39&36\\
$[45,50)$&23&16\\
\geq 50&12&12\\
 \end{array}
$$
\end{table}

\begin{table}
\caption{Interval, observed and expected void frequencies of the data set, $R^{(3)}_0$ of N-body Mock Catalog with the estimates $\lnm(0.223, 3.412, 0.301)$}
$$\begin{array}{l|r|r|r}\text{Interval}&\text{Observed\phantom{a}Frequencies} & \text{Expected\phantom{a}Frequencies}\\\hline
$[0,5)$&0&0\\
$[0,20)$&11\,487&11\,495\\
$[20,25)$&28\,035&25\,508\\
$[25,30)$&31\,631&33\,088\\
$[30,35)$&27\,770&29\,407\\
$[35,40)$&20\,605&20\,781\\
$[40,45)$&12\,894&12\,718\\
$[45,50)$&7\,295&7\,098\\
$[50,55)$&3\,848&3\,732\\
$[65,70)$&2\,129&1\,889\\
\geq 60&1\,834&1\,812\\
 \end{array}
$$
\end{table}
\noindent
Second, graphically by Figure 3 which displays the sample histograms with the curves of the log-normal densities whose parameters are the estimates computed from the samples. In both cases the goodness-of-fit is evidently fairly good. Note that for $R^{(1)}$ the p-value associated with Kolmogorov, Cram\'er-von Mises and Anderson-Darling statistics equals the remarkably high value of $0.99$.

\begin{figure*}
\begin{tabular}{ll}
\includegraphics[scale=0.55]{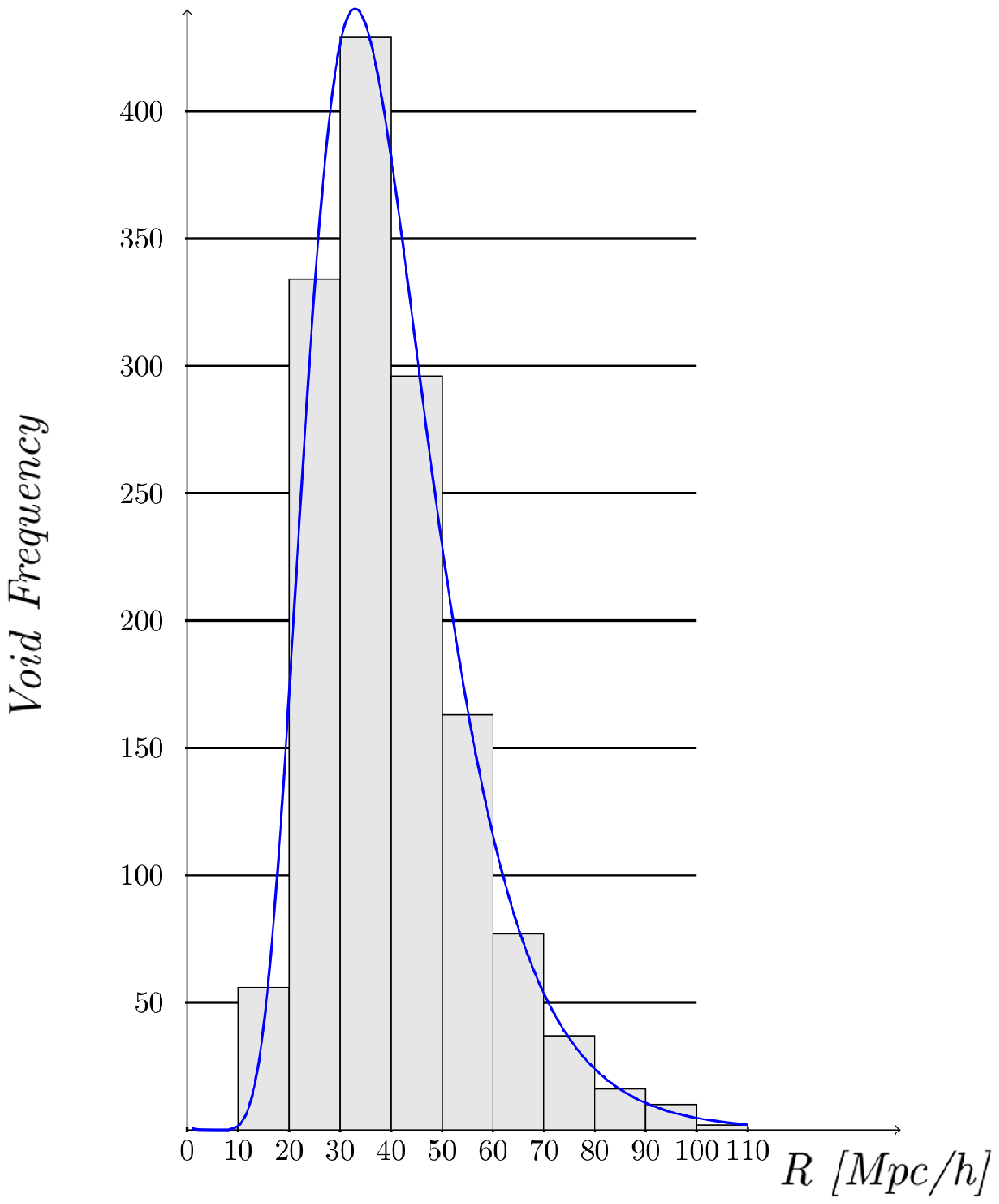}
\includegraphics[scale=0.6]{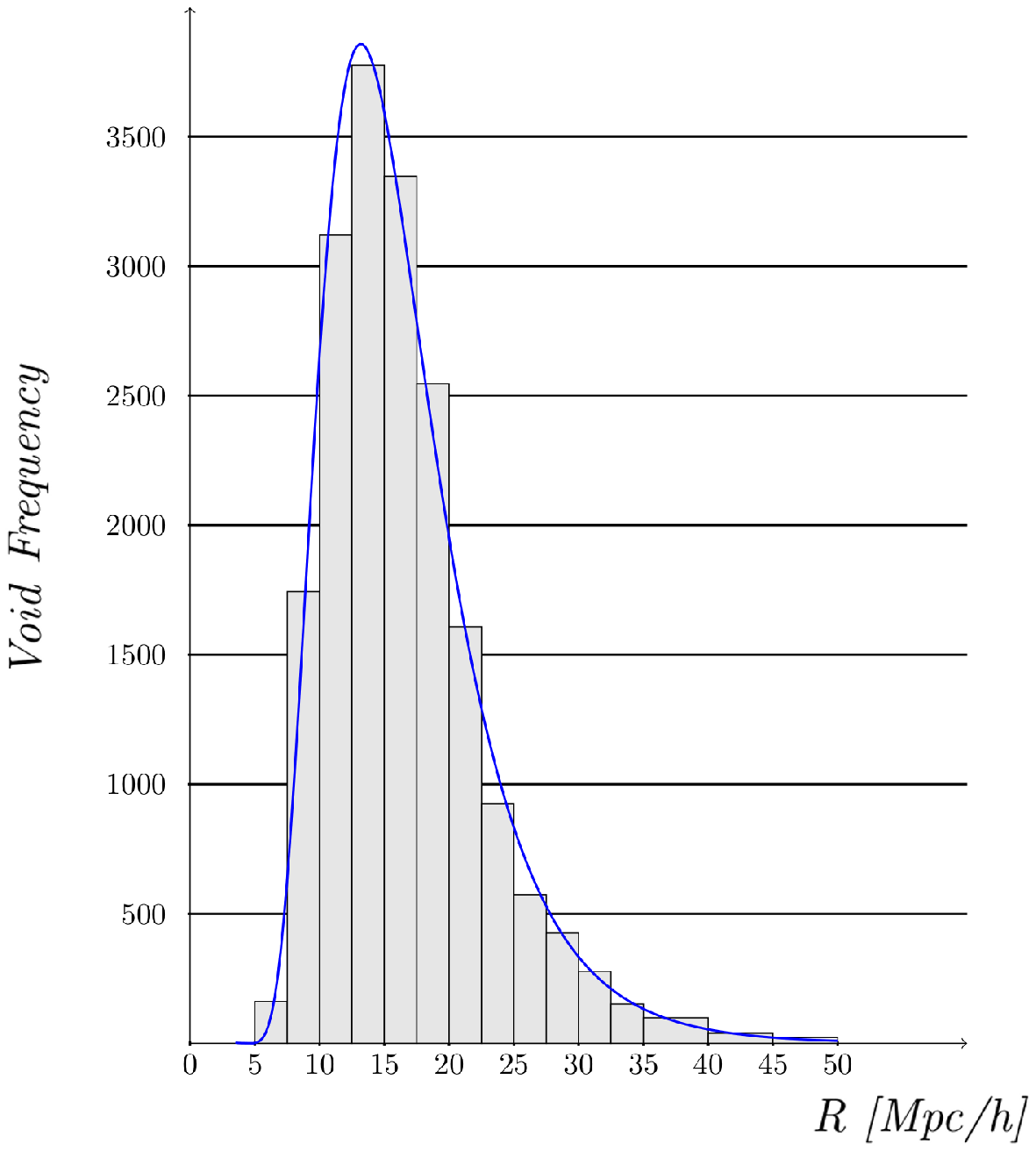}\\
\includegraphics[scale=0.9]{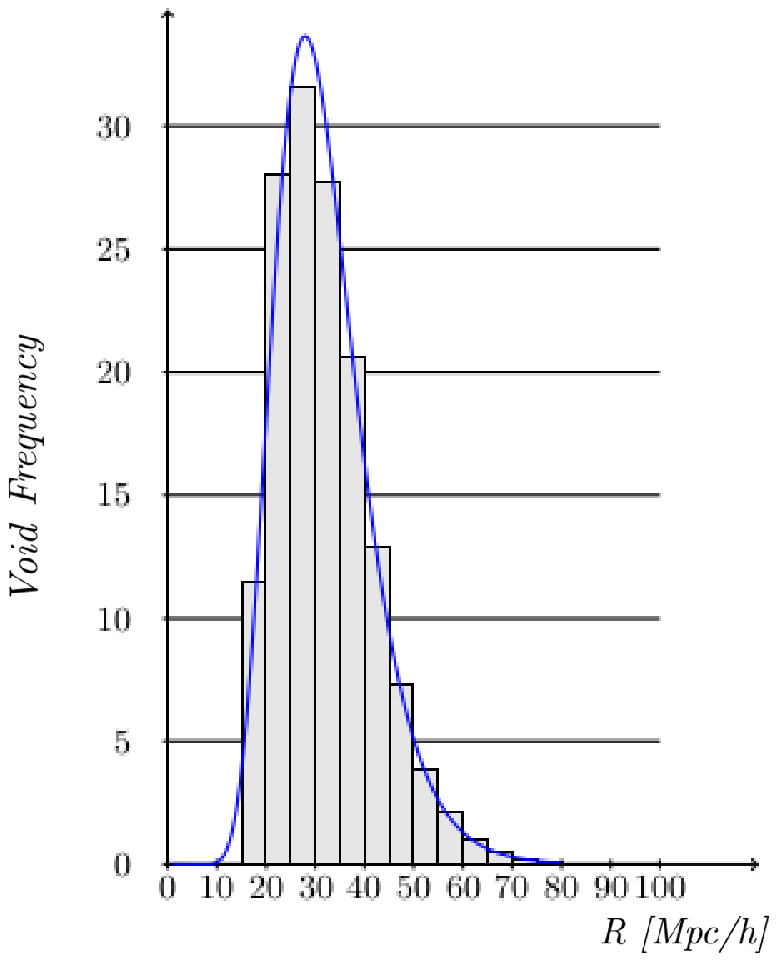}
\end{tabular}
\caption{Void frequencies from the samples $R^{(1)}$, $R^{(2)}$ and $R^{(3)}_{0}$ with the density curves of $\lnm(0.770, 3.610, 0.373)$, $\lnm(3.284, 2.493, 0.449)$, $\lnm(0.223,3.412,0.301)$, respectively (from top left, top right and bottom left).}
\end{figure*}
\noindent
\section{Conclusions and Discussion}

Today we know that voids dominate the total observed volume of the large scale structure \citep{kir,gehu,dac,shect} and they are very sensitive to their environments which can strongly affect their shape as well their distributions \citep{sw,Esra1,Esra2}. Therefore the void size distribution functions may play an important role to understand the dynamical processes affecting the structure formation of the Universe \citep{Croton2005,goldbergvogeley,Hoyle2005}.

In this study, we show that the system of $3$-parameter log-normal distributions gives a satisfactory model of the size distribution of voids obtained from the three different HOD mock catalogs of CVC; N-body Mock, HOD Sparse and HOD Dense. These catalogs are especially important since they may allow us to compare HOD mock voids in different environments. As aforementioned, HOD Dense and HOD Sparse provide us the data sets to obtain the size distributions at $z=0$ \citep{sutter2014a}, while N-body Mock catalog has $155,196$ voids at $z=0.53$ in the full volume-set of mock galaxies \citep{sutter2014b}. We find that N-body Mock catalog, $R^{(3)}$, show sub-void populations with respect to their central densities (Left Panel, Figure 1). From these sub-populations we choose the population includes voids with only zero central density. This sub-sample is named as $R^{(3)}_{0}$ which consists of $147,528$ voids and the maximum tree depth $0$. The size distribution obtained from this sub-sample satisfies the $3$-parameter log-normal distribution as other catalogs. As is seen, this sub-sample has the smallest value of the maximum tree depth compared to the two catalogs we use in this study. This is especially important since it provides us a third environment to compare the relationship between the maximum tree depth and the two shape parameters of the log-normal size distribution. Then, we compare the void size distributions obtained from these mock data. As a result, we show that all three samples fit the $3$-parameter log-normal distribution. As is seen in Fig 2 and Table 3, the HOD Sparse sample, $R^{(1)}$ remarkably passes the Kolmogorov, Cram\'er-von Mises and Anderson-Darling statistical test with high p-value of $0.99$, which seldom occurs with a large size sample.

Recalling that skewness $b_{1}$ of the log-normal distribution indicates the asymmetry of the void size distribution, in Fig 2, Table 3, 4 and 5, HOD Dense $R^{(2)}$, is the most asymmetric sample with the skewness $b_{1}=2.322$ compared to the void populations in N-body Mock, $R^{(3)}_{0}$ with $b_{1}=0.912$ and HOD Sparse, $R^{(1)}$ with $b_{1}=1.488$ (also see Table 1 to compare other parameters of the void size distributions between the three samples). Also, the void size distribution of N-body Mock $R^{(3)}_{0}$ is the least asymmetric of the three samples. In addition, we find that the void distribution in HOD Dense indicates higher kurtosis, $b_{2}\approx7$ than the void population in the HOD Sparse sample, $b_{2}\approx 5.5$, while the N-body Mock sub-sample shows the lowest kurtosis $b_{2}=4.134$. As is seen that there may be a relation between the values of kurtosis, skewness and the maximum tree depth of the void size distribution. Note that HOD Dense has the highest maximum tree depth $10$ as well as the highest values of the kurtosis $b_{2}\approx7$ and skewness $b_{1}=2.322$ of the three samples, while the size distribution obtained from the N-body Mock sub-sample $R^{(3)}_{0}$ consists of the lowest values of the distribution shape parameters ($b_{1}=0.912$ and $b_{2}=4.134$) and the maximum tree depth $0$ taking into account only the root voids. Considering that the maximum tree depth is a measure of the amount of void substructures, it may be possible to determine a connection between the number of void substructures and how the void size distribution is skewed and peaked.

Note that all the void size distributions have positive skewness and the strength of the skewness may be related to the length of the tail of the distribution in radius (see Figure 3). In addition, it is natural to see a thin tail due to the shape of the log-normal distribution in any case. For example, HOD Dense ($R^{(2)}$) and HOD Sparse ($R^{(1)}$) show slightly longer tails on the large radius side compared to the distribution of the N-body Mock sub-sample ($R^{(3)}_{0}$) which has the smallest skewness (Table 1 and Figure 3). These thin tail formations of the log-normal void size distributions derived from the simulations are particularly interesting due to a possible connection of dynamical interplay between void size and galaxy distributions. Assuming the void size is proportional to its mass content ($R^3\approx M$), we here realized that the one-point galaxy distribution obtained by \cite{Kayo} from the N-body simulations seems strikingly similar to the log-normal void distribution that we obtain (see equation (\ref{eqn:void-distribution})). On the other hand, the void log-normal distribution has one more additional parameter than the galaxy distribution of \cite{Kayo} which provides a better fit to the void data sets. From small to large scale matter distributions in the Universe, in other words from the interstellar to the intergalactic medium, there are models and observational studies to explain the skewed log-normal matter distribution. For example, \cite{Schneider2013} conclude that statistical density fluctuations, intermittency, and magnetic fields can cause excess from the log-normal distribution, and also that core formation and/or global collapse of filaments and a non-isothermal gas distribution lead to a power-law tail. \cite{pudritz} show that the observed skewed log-normal galaxy distribution could arise from repeated shock interactions, without the need for fully-developed turbulence. Recently it has been shown that it is possible to see large scale bulk flows which indicate that the possibility of the effects of large scale turbulence on structure formation may not be ignored. Related to this, \cite{wang12} find a large-scale bulk flow of approximately a sphere of radius $170$ $Mpc/h$ produced by the massive structures associated with the SDSS Great Wall. Taking into account that voids and galaxies/overdensities are formed from the same primordial density field, then the similarity between the void size and galaxy mass distributions may be expected. As a result, it is possible that the galaxy and void distributions are analogous to each other. This analogy may be caused by void substructures due to the inner as well as outer tidal streams since it is shown that the interiors of voids are filled with subvoids, galaxies and even filaments by $N$-body simulations and observations \citep{gott,mathis,benson,Patiri2006MNRAS.372.1710P,Kreckel2011}. Note that proving this connection between void and galaxy mass distribution deserves its own systematic statistical study, especially by using real data sets, therefore we leave the answer of this puzzle to a future study.

In addition to this, the peaks of the distributions is given by $\overline r$ in Table 1. $\overline r$ provides the most dominant void sizes in the void size distribution. Figure 3 and Tables 1 show that $\approx 40.4$ $Mpc/h$ size voids dominate the HOD Sparse sample, $R^{(1)}$, while the size distributions of the HOD Dense $R^{(2)}$ and the N-body Mock $R^{(3)}_{0}$ are dominated by voids with radii $\approx 17$ $Mpc/h$ and $\approx 32$ $Mpc/h$ respectively. This indicates that the size distributions of HOD Sparse and the N-body Mock sub-sample tends to have larger voids compared to the voids in HOD Dense due to their low density environment. This result agrees with \cite{Watson2014,Jennings2013,Esra1,Esra2}. On the other hand, as aforementioned, the N-body Mock catalog has higher redshift $z=0.53$ than the two other samples, therefore N-body Mock sample may imply larger size voids at $z=0$ compared to HOD Sparse.

Although there are some theoretical attempts to understand the dynamical, thermal and chemical evolution of the void population, and the interplay between galaxies and the intergalactic medium \citep{Viel2008,Shang2007,sw,Esra1,Esra2}, a systematic investigation of void size distributions is still missing. Therefore, in this study we show that the void size distributions satisfy the $3$-parameter log-normal distribution and this distribution may be a good candidate to show the large scale environmental effects on voids. It also seems that there is a relation between the strength of the maximum tree depth and the $3$-parameter log-normal void size distribution parameters; skewness and kurtosis. On the other hand, an extended statistical study of the size distribution of voids in real as well as more simulated data is crucial to fully understand the effects of the large scale dynamical network between galaxies, filaments and voids to obtain more precise results.

\acknowledgments
The authors would like to thank Paul Sutter for insightful comments and suggestions on the Cosmic Void Catalogs. The three void catalogs used here can be found in folder 


\begin{thebibliography}{}
\bibitem[Benson et al.(2003)]{benson} Benson, A.~J., Hoyle, F., Torres, F., \& Vogeley, M.~S., 1986, \mnras, 340, 160
\bibitem[Bernardeau(1992)]{Bernardeau92} Bernardeau, F., 1992, \apj, 392, 1
\bibitem[Bernardeau(1994)]{Bernardeau94} Bernardeau, F., 1994, \aap, 291, 697
\bibitem[Bouchet et al.(1993)]{Bouchet1993} Bouchet, F.~R., et al., 1993, \apj, 417, 36
\bibitem[Coles \& Jones(1991)]{colesjones91} Coles, P. \& Jones, B., 1991, \mnras, 248, 1
\bibitem[Colombi et al.(1995)]{ColombiBS1995} Colombi, S., Bouchet, F.~R., \& Schaeffer, R., 1995, \apjs, 96, 401
\bibitem[Croton et al.(2005)]{Croton2005} Croton, D. J. et al., 2005, \mnras, 356, 1155
\bibitem[da Costa et al.(1994)]{dac} da Costa, L.~N. et al., 1994, \apjl, 424, L1
\bibitem[Dawson et al.(2013)]{Dawson2013} Downson, K.~S., 2013, \mnras, 145, 10
\bibitem[Elizalde \& Gaztanaga(1992)]{elizaldegaztanaga92} Elizalde, E., \& Gaztanaga, E., 1992, \mnras, 254, 247
\bibitem[Fisher(1993)]{fisher} Fisher, N. I., 1993, Statistical analysis of circular data (Cambridge, Cambridge University Press)
\bibitem[Fry(1986)]{fry86Voids} Fry, J.~N., 1986, \apj, 306, 358
\bibitem[Gazta{\~{n}}aga \& Yokoyama(1993)]{Gaztanaga93} Gazta{\~{n}}aga, E., \& Yokoyama, J., 1993, \apj, 403, 450
\bibitem[Geller \& Huchra(1989)]{gehu} Geller, M.~J., \& Huchra, J.~P., 1989, Science, 246, 897
\bibitem[Goldberg \& Vogeley(2004)]{goldbergvogeley} Goldberg, D.~M., \& Vogeley, M.~S., 2004, \apj, 605, 1
\bibitem[Gottl{\"o}ber et al.(2003)]{gott} Gottl{\"o}ber, S., {\L}okas, E.~L., Klypin, A., \& Hoffman, Y.,  2003, \mnras, 344, 715
\bibitem[Hamilton(1985)]{Hamilton1985} Hamilton, A.~J.~S., 1985, \apjl, 292, L35
\bibitem[Hoyle et al.(2005)]{Hoyle2005} Hoyle, F., et al., 2005, \apj, 620, 618
\bibitem[Jennings et al.(2013)]{Jennings2013} Jennings, E., Li, Y., \& Hu, W., 2013, \mnras, 434, 2167
\bibitem[Johnson et al.(1994)]{kotz} Johnson, N.L, Kotz, S, \& Balakrishnan, N., 1994, Continuous Univariate Distributions, Vol. 1 (2nd ed; Wiley)
\bibitem[Kayo et al.(2001)]{Kayo} Kayo, I., Taruya, A., \& Suto, Y., 2001, \apj, 561, 22
\bibitem[Kendall \& Stuart(1977)]{stuart1} Kendall, M., \& Stuart, A., 1977, The advanced theory of statistics. Vol 1: Distribution Theory, 2nd Ed., (New York, NY., Macmillan)
\bibitem[Kendall \& Stuart(1977)]{stuart3} Kendall, M., \& Stuart, A., 1977, The advanced theory of statistics. Vol 3: Design and Analysis, and Time-series, 2nd Ed., (New York, NY., Macmillan)
\bibitem[Kirshner et al.(1981)]{kir} Kirshner, R.~P., Oemler, Jr., A., Schechter, P.~L., \& Shectman, S.~A., 1981, \apjl, 248, L57
\bibitem[Kofman et al.(1994)]{Kofman} Kofman, L., Bertschinger, E., Gelb, J.~M., Nusser, A., \& Dekel, A., 1994, \apj, 420, 44
\bibitem[Kreckel et al.(2011)]{Kreckel2011} Kreckel, K., et al. 2011, \aj, 141, 4
\bibitem[Lahav et al.(1993)]{Lahav93} Lahav, O., Itoh, M., Inagaki, S., \& Suto, Y., 1993, \apj, 402, 387
\bibitem[Lavaux \& Wandelt (2012)]{Lavaux12} Lavaux, G., \& Wandelt, B.~D., 2012, \apj, 754, 109
\bibitem[Lee \& Hoyle(2015)]{mathis} Lee, J., \&  Hoyle, F., 2015, \apj, 803, 45
\bibitem[Manera et al.(2013)]{Manera2013} Manera, M., et al., 2013, \mnras, 428, 1036
\bibitem[Mathis \& White(2002)]{mathis} Mathis, H., \&  White, S.~D.~M., 2002, \mnras, 337, 1193
\bibitem[Neyrick(2008)]{Neyrinck2008} Neyrinck, M.~C., 2008, \mnras, 386, 2101
\bibitem[Patiri et al.(2006a)]{patiri2006a} Patiri, S.~G., Betancort-Rijo, J., \& Prada, F., 2006, \mnras, 368, 1132
\bibitem[Patiri et al.(2006b)]{Patiri2006MNRAS.372.1710P} Patiri, S.~G., Prada, F., Holtzman, J., Klypin, A. \& Betancort-Rijo, J., 2006, \mnras, 372, 1710
\bibitem[Planck Collaboration(2014)]{PlanckCollaboration} Planck Collaboration, 2014, \aap, 571, A19
\bibitem[Pudritz \& Kevlahan(2013)]{pudritz} Pudritz, R.~E., \& Kevlahan, N.~K.-R., 2013, Philosophical Transactions of the Royal Society of London Series A, 371, 20248
\bibitem[Russell(2013)]{Esra1}  Russell, E., 2013, \mnras, 436, 3525
\bibitem[Russell(2014)]{Esra2} Russell, E., 2014, \mnras, 438, 1630
\bibitem[Ryu et al.(2003)]{Ryu} Ryu, D., Kang, H., Hallman, E., \& {Jones}, T.~W., 2003, \apj, 593, 599
\bibitem[Saslaw(1985)]{saslaw} Saslaw, W.~C., 1985, Gravitational physics of stellar and galactic systems, Cambridge, 506, Cambridge University Press
\bibitem[Schneider et al.(2013)]{Schneider2013} Schneider, N. et al., 2013, \apjl, 766, L17
\bibitem[Shang et al.(2007)]{Shang2007} Shang, C., Crotts, A., \& Haiman, Z., 2007, \apj, 671, 136
\bibitem[Shectman et al.(1996)]{shect} Shectman, S.~A. et al., 1996, \apj, 470, 172
\bibitem[Sheth \& van de Weygaert (2004)]{sw} Sheth, R.~K., \& van de Weygaert, R., 2004, \mnras, 350, 517
\bibitem[Sheskin(2011)]{sheskin} Sheskin, D. J., 2011, Handbook of parametric and nonparametric statistical procedures (Boca Raton: FL, CRC Press)
\bibitem[Strauss et al. (2002)]{Strauss2002} Strauss, M.~A., 2002, \aj, 124, 1810
\bibitem[Sutter et al.(2012)]{sutter} Sutter, P.~M., Lavaux, G., Wandelt, B.~D., \& Weinberg, D.~H., 2012, \apj, 761, 44
\bibitem[Sutter et al.(2014a)]{sutter2014a} Sutter, P.~M., Lavaux, G., Hamaus, N., Wandelt, B.~D., Weinberg, D.~H. \& Warren, M.~S., 2014, \mnras, 442, 462
\bibitem[Sutter et al.(2014b)]{sutter2014b} Sutter, P.~M., Lavaux, G., Wandelt, B.~D., Weinberg, D.~H., Warren, M.~S., \& Pisani, A., 2014, \mnras, 442, 3127
\bibitem[Szapudi\& Colombi(1996)]{SzapudiC1996} Szapudi, I., \& Colombi, S., 1996, \apj, 470, 131
\bibitem[Taylor \& Watts (2000)]{TaylorWatts2000} Taylor, A.~N., \& Watts, P.~I.~R., 2000, \mnras, 314, 92
\bibitem[Tinker et al.(2006)]{Tinker2006} Tinker, J.~L., Weinberg, D.~H., \& Zheng, Z., 2006, \mnras, 368, 85
\bibitem[Ueda \& Yokoyama (1996)]{Ueda} Ueda, H., \& Yokoyama, J., 1996, \mnras, 280, 754
\bibitem[Viel et al.(2008)]{Viel2008} Viel, M., Colberg, J.~M., \& Kim, T.-S., 2008, \mnras, 386, 1285
\bibitem[Wang et al.(2012)]{wang12} Wang, H., Mo, H.~J., Yang, X., \& van den Bosch, F.~C., 1979, \mnras, 420, 1809
\bibitem[Watson et al.(2014)]{Watson2014} Watson, W.~A. et al., 2014, \mnras, 437, 3776
\bibitem[White(1979)]{white79} White, S.~D.~M. 1979, \mnras, 186, 145
\bibitem[Zehavi et al.(2011)]{Zehavi2011} Zehavi, I. et al., 2011, \apj, 736, 59
\bibitem[Zheng et al.(2007)]{Zheng07} Zheng, Z., Coil, A.~L., \& Zehavi, I., 2007, \apj, 667, 760
\end{thebibliography}
\end{document}